\documentclass[twocolumn,showpacs,preprintnumbers,prl,amsmath,amssymb, letterpaper]{revtex4}
\usepackage{verbatim,graphicx} 
\usepackage{dcolumn}  
\usepackage{bm}

\newcommand{\be}{\begin{equation}}
\newcommand{\ee}{\end{equation}}
\newcommand{\lb}{\label}

\newcommand{\br}{{\bf r}}

\newcommand{\bx}{{\bf x}}

\newcommand{\bI}{{\bf I}}

\newcommand{\bS}{{\bf S}}

\newcommand{\bX}{{\bf X}}

\newcommand{\BR}{{\Bbb R}}

\newcommand{\boalpha}{{\mbox{\boldmath $\alpha$}}}

\newcommand{\bomu}{{\mbox{\boldmath $\mu$}}}

\newcommand{\boxi}{{\mbox{\boldmath $\xi$}}}

\newcommand{\bzed}{{\mbox{\boldmath $0$}}}

\newcommand{\beg}{\begin{equation}}
\newcommand{\eeq}{\end{equation}}

\begin{document}

\title{Equation-Free Implementation of Statistical Moment Closures}

\author{ Francis J. Alexander and Gregory Johnson}
\affiliation{Los Alamos National Laboratory, P.O.Box 1663, \\
Los Alamos, NM, 87545.}
\author{   Gregory L. Eyink}
\affiliation { {\em Department of Mathematical Sciences} \\ 
 {\em Johns Hopkins University}\\ 
  {\em Baltimore, MD 21218}}
\author{Ioannis G. Kevrekidis}
\affiliation{  {\em Department of Chemical Engineering and PACM} \\ 
  {\em Princeton University} \\
  {\em Princeton, NJ 08544}}

\begin{abstract}

We present a general numerical scheme for the practical implementation
of statistical moment closures suitable for modeling complex,
large-scale, nonlinear systems. Building on recently developed
equation-free methods, this approach numerically integrates the
closure dynamics, the equations of which may not even be available in
closed form.  Although closure dynamics introduce statistical
assumptions of unknown validity, they can have significant
computational advantages as they typically have fewer degrees of
freedom and may be much less stiff than the original detailed
model. The closure method can in principle be applied to a wide class
of nonlinear problems, including strongly-coupled systems (either
deterministic or stochastic) for which there may be no scale
separation. We demonstrate the equation-free approach for implementing
entropy-based Eyink-Levermore closures on a nonlinear stochastic
partial differential equation.

\end{abstract}
\maketitle

\section{Introduction}
Accurate, fast simulations of complex, large-scale, nonlinear systems
remain a challenge for computational science and engineering, despite
extraordinary advances in computing power.  Examples range from
molecular dynamics simulations of proteins~\cite{MDProteinsI},
\cite{MDProteinsII} and glasses~\cite{MDGlasses}, to stochastic
simulations of cellular biochemistry ~\cite{Gillespie,Wilkinson}, 
to global-scale, geophysical fluid dynamics~\cite{MajdaWang}.  
Often for the systems under consideration there is no 
obvious scale separation, and their many degrees of freedom are
strongly coupled.  The complex and multiscale nature of these
processes therefore makes them extremely difficult to model
numerically.  To make matters worse, one is often interested not in a
single, time-dependent solution of the equations governing these
processes, but rather in ensembles of solutions consisting of multiple
realizations (e.g., sampling noise, initial conditions, and/or
uncertain parameters).  Often real-time answers are needed (e.g., for
control, tracking, filtering).  These demands can easily exceed the
computational resources available not only now but also for the
foreseeable future.

In principle, all statistical information for the problem under
investigation is contained in solutions to the Liouville (if
deterministic)/Kolmogorov (if stochastic) equations.  These are
partial differential equations in a state space of high (possibly
infinite) dimension.  A straightforward discretization of the
Liouville / Kolmogorov equations is therefore impractical.  An
ensemble approach to solving these equations can be
taken; however, quite often, the practical application of the ensemble
approach is also problematic. Generating a sufficient number of
independent samples for statistical convergence can be a challenge.
For some problems, computing even {\em one} realization may be
prohibitive.

The traditional approach to making these problems computationally
tractable is to replace the Liouville/Kolmogorov equation by a (small)
set of equations (PDEs or ODEs) for a few, low order statistical
moments of its solution. When taking this approach for nonlinear
systems, one must make an approximation, a closure, for the dependence
of higher order moments on lower order moments.  Typically the form of
the closure equation is based on expert knowledge, empirical data,
and/or physical insight.  For example, in the superposition
approximation and its extensions~\cite{Superposition} for dense
liquids and plasmas, both quantum or classical, one approximates third
order moments as functions of second order moments.  Moment closure
methods of this type have been applied to a number of areas including
fluid turbulence (see ~\cite{TurbClosure} and references therein, and
also the work of Chorin {\em et. al.}). Of course, as with any
approximation strategy, the quality of the resulting reduced
description depends on the approximations made -- poor closures lead
to poor answers/predictions.  In addition to replacing the ensemble
with a small set of equations for low order moments, these equations
are typically easier to solve. They are deterministic and generally
far less stiff than the original equations.

A less exploited variant of this approximation scheme is the probability density
function (PDF) based moment-closure approach. For PDF moment closures
one makes an {\em ansatz} for the system statistics guided by available 
information (e.g., symmetries).  One then uses this {\em ansatz} in conjunction 
with the original dynamical equations to derive moment equations. Such 
PDF-based closures have been developed for reacting scalars advected 
by turbulence \cite{CCK}, phase-ordering dynamics \cite{Oono} and a variety 
of other systems. This approach to moment-closure is a close analogue 
of the Rayleigh-Ritz method frequently used in solving the quantum-mechanical 
Schroedinger equation, by exploiting an {\it ansatz} for the wave-function. 
For a formal development of this point of view, see  \cite{Eyink96}.

One of the obstacles to applying moment closures is that often the
closure equations are too complicated to write down explicitly, even
with the availability of computer algebra / symbolic computation
systems.  This is especially true for large-scale, complex systems,
e.g. global climate models.  Because of their great complexity, even
if one could in principle derive the closure equations analytically,
this procedure would be extremely difficult and time-intensive.
Moreover, each time a model is updated, as climate and ocean models
regularly are, the closure equations would have to be rederived.  In
other cases it may simply be impossible to determine the closure
equations analytically. This is especially likely when PDF's are not
Gaussian, which is the case for most {\em useful} closures.  Monte
Carlo or other numerical methods may be needed in order to evaluate
integrals for the moments~\cite{Levermore96}.  In addition, there may
be situations where neither analytic nor numerical/MC integration will
yield the closure equations due to the black-box nature of the
available numerical simulator such as a compiled numerical code with
an inaccessible source. Clearly, a need exists for a robust approach
to the general closure protocol which circumvents analytical
difficulties.

We address that need here by combining PDF closures with equation
free modeling~\cite{EQNFREE}~\cite{EqnFree2}.  The basic premise of
the equation-free method is to use 
an ensemble of short bursts of simulation of the
original dynamical system 
to estimate, on demand, the time-evolution of the
the closure equations that we may not explicitly have.  The
equation-free approach extends the applicability of statistical
closures beyond the rare cases where they can be expressed in closed
form. This hybrid strategy may be faster than the brute-force solution of
a large ensemble of realizations of the dynamical equations since
the closure version is generally smoother than the original problem.

This paper is organized as follows. In Section 2 we describe the
general features of PDF-based moment closures.  In Section 3 we
explain how to implement the equation-free approach with these
closures.  We then, in Section 4, apply these ideas for a specific
dynamical system, the stochastic Ginzburg-Landau (GL) equations 
using a particular PDF-based closure scheme, the entropy method 
of Eyink and Levermore~\cite{EL}.  We conclude with a discussion 
of closure quality, computational issues, and the application of our
approach to large-scale systems.

\section{PDF-Based Moment Closures}

We consider the very general class of dynamical systems, including
maps, formally represented by
\begin{eqnarray}
{\dot {\bf X}} = {\bf U}({\bf X}(t),{\bf N}(t),t)
\end{eqnarray}
{\rm or}
\begin{eqnarray}
{{\bf X}_{t+1}} = {\bf U}_t({\bf X}_t,{\bf N}_t)
\label{eq:genform}
\end{eqnarray}
where ${\bf N}(t)$ is a stochastic process with prescribed statistics.  
The stochastic component arises from unknown parameters, random forcing, 
neglected degrees of freedom and/or random initial conditions.  This class 
includes both deterministic and stochastic systems with discrete and/or 
continuous states. Queueing systems, molecular dynamics, and stochastic PDEs 
are just some of the many examples that fall into this category.

For concreteness in this paper we restrict ourselves to a special case
of equation~(\ref{eq:genform}), namely, situations where ${\bf N}(t)$ is a 
Markov process (Brownian motion, Poisson process, etc.) and---more
specifically still---It\^{o} stochastic differential equations of the form:
\begin{equation} 
d{\bf{X}} = {\bf U}({\bf{X}},t) dt + \sqrt{2} {\bf S} ({\bf X},t) d {\bf
W}(t)
\label{eq:Dynamics}.
\end{equation} 
The deterministic component of the state, ${ \bf X }$, is governed by
the continuously differentiable vector field, ${\bf U}:\BR^N \times \BR
\to \BR^N$.  For many problems of interest (e.g., climate) ${\bf U}$ is
a highly nonlinear function.  The noise component is modeled by the
standard mean $\bzed$, covariance matrix $\bI$ Wiener process, ${\bf W}\in\BR^N$, 
possibly modulated by a state-dependent matrix ${\bf S} : \BR^N \times \BR 
\to \BR^{N\times N}$.  Equation (\ref{eq:Dynamics}) encompasses a wide 
class of systems including deterministic $({\bf S} =0)$ ones.

In many cases one is interested in knowing the low order statistics of
equation~(\ref{eq:Dynamics}), for example an instantaneous mean value
or possibly multi-point covariance of ${\bf X}$.  These statistics can
be obtained by averaging over an ensemble of stochastic systems,
solving equation~(\ref{eq:Dynamics}).  They can also be obtained via 
the forward Kolmogorov equation for the probability density function 
$P({\bf X},t)$:
\begin{equation}
\partial_t P = {\cal{L}}^* (t) P,
\label{eq:Kolmogorov}
\end{equation}
where  $P$ satisfies the conditions: $P({\bf
X},t) \ge 0$, and $\int P ({\bf X}, t) \,d{\bf X} = 1$,
and where ${\cal L}^*$ is the generator of the Markov process.
In the case of equation~(\ref{eq:Dynamics}) this operator takes the 
form
\begin{equation}
{\cal{L}}^*(t) \psi({\bf X})= - \nabla_{{\bf X}} \cdot ({\bf U}({\bf X},t)\psi
({\bf X})) + \nabla_{{\bf X}}^2 :({\bf D}({\bf X},t) \psi ({\bf X})).
\end{equation}
The forward Kolmogorov equation then becomes a Fokker-Planck equation
\begin{equation} \partial_t P+ \nabla_{{\bf X}} \cdot ({\bf U}P) = 
\nabla_{{\bf X}}^2 :({\bf D}P)
\end{equation}
where ${\bf D}({\bf X},t)=\bS({\bf X},t)\bS({\bf X},t)^T$ is the nonnegative-definite
diffusion matrix arising from the noise term.
Unlike the original dynamical equation~(\ref{eq:Dynamics}), the
forward Kolmorogov equation (FKE) is both linear and deterministic.
Dealing with it, therefore, has apparent advantages over the 
original ensemble of stochastic systems simulations.  The price to
pay for these advantages is that the FKE lives in a typically high,
potentially infinite-dimensional, space.  When equation~(\ref{eq:Dynamics}) 
is a nonlinear PDE, numerical solution to the FKE is usually ruled out.

For computational purposes, we would therefore like to reduce the FKE
(if possible and useful) to a small system of ordinary differential
equations.  This reduction should simplify the computation as much as
possible while retaining fidelity to the original dynamical processes.
The reduction proceeds by taking moments of the FKE with respect to a
vector-valued function $\boxi ({\bf X},t) $ from $ \BR^N \times \BR_+
\to \BR^M$.  The $\boxi$ selected should include the relevant
variables in the system (slow modes, conserved quantities, etc.).  The
moments $\bomu(t)$ of $\boxi({\bf X},t)$ are defined by
\begin{eqnarray}
\bomu (t) = \int \boxi ({\bf X},t)P({\bf X},t)d{\bf X}
\end{eqnarray}
and give rise to 
\begin{eqnarray}
{\dot \bomu(t)} = \int {\dot \boxi} ({\bf X},t)P({\bf X},t)d{\bf
X}, \label{mudot}
\end{eqnarray}
where
\begin{eqnarray}
{\dot \boxi}({\bf X},t) = \partial_t \boxi({\bf X},t) + {\cal L}(t)
\boxi ({\bf X},t)
\end{eqnarray}
and ${\cal L}$ is the adjoint of ${\cal L}^*$ or the backward Kolmogorov 
operator. The result (\ref{mudot}) can be obtained by averaging over an 
ensemble of realizations of the stochastic dynamics~(\ref{eq:Dynamics}). 
In general, however, (\ref{mudot}) is not a closed equation for the 
moments, $\bomu$.  One can close this equation by choosing a PDF, 
$P({\bf X}, t,{\bomu})$, which itself is a function of the moments $\bomu$.
\begin{eqnarray}
{\dot \bomu} (t) = {\bf V}(\bomu,t) \equiv \int {\dot \boxi} ({\bf
X},t)P({\bf X},t,\bomu)d{\bf X}.
\label{eq:Veq}
\end{eqnarray}
Alternatively, one can select a family of probability densities
$P({\bf X},t, \boalpha)$, specified by parameters $\boalpha =
\boalpha(\bomu,t)$ rather than directly by the moments ${\bomu}$.  This is
analogous to specifying the temperature in the canonical ensemble as opposed
to the average energy.  The
equivalence of these approaches is guaranteed provided that the
parameters and moments can be determined uniquely from one another.
The translation between the parameters and their corresponding moments
can be carried out by one of several methods. In some cases one may
require Monte Carlo evaluation of the resulting integrals.  

If the moments and/or parameters are selected judiciously, one hopes that 
the approximate PDF $P({\bf X},t, \boalpha ( \bomu)(t) )$ will be close to the 
exact solution of the Liouville/Kolmogorov equation (\ref{eq:Kolmogorov}).
The mapping closure approach of Chen et al~\cite{CCK} and the Gaussian
mapping method of Yeung et al.~\cite{Oono} are based on this type of
parametric PDF closure~\footnote{In the case of ~\cite{CCK} the dynamics
is an advection-reaction-diffusion equation for a scalar concentration field 
$\bX(t)=\{\theta({\bf x},t):\,\bx\in\BR^d\}$. The moment functions are the 
``fine-grained PDF'' $\xi_{\vartheta,{\bf x}}[\bX,t]=\delta(\theta({\bf x},t)-\vartheta),$
labelled by space point ${\bf x}$ and scalar value $\vartheta.$ The
moment average $\mu_{\vartheta,{\bf x}}(t)=\langle\delta(\theta({\bf x},t)
-\vartheta)\rangle$ is the 1-point PDF $p(\vartheta;\bx,t)$ which gives
the distribution of scalar values $\vartheta$ at space-time point $({\bf x},t).$ 
The parametric model $P[\bX;\boalpha,t]$ is the distribution over scalar fields 
obtained by the {\it ansatz} $\theta({\bf x},t)=X(\theta_0({\bf x},t),{\bf x},t)$ 
where $\theta_0({\bf x},t)$ is a reference random field of known (Gaussian) 
statistics and $X(\cdot,{\bf x},t): \BR \rightarrow \BR$ is a ``mapping function". 
The latter function is the ``parameter'' $\alpha_{\vartheta_0,{\bf x}}(t)=
X(\vartheta_0,{\bf x},t)$ which determines (and is determined by) the 
``moment'' $\mu_{\vartheta,{\bf x}}(t)$ from the relation 
$p(X(\vartheta_0,\bx,t);\bx,t)|\partial X/\partial \vartheta_0|=
p_0(\vartheta_0,\bx,t).$ Here $p_0$ is the 1-point PDF of the 
reference Gaussian field $\theta_0(\bx,t).$ \\
\vspace{4pt} $\,\,\,\,\,\,\,$The approach of \cite{Oono} is similar. The 
problem is phase-ordering dynamics as given, for example, by our
equation (\ref{eq:TDGL}) and $\bX(t)=\{\phi({\bf x},t):\,\bx\in\BR^d\}.$
The moment functions are the quadratic products $\xi_\br[\bX,t]
=\phi(\br,t)\phi(\bzed,t),$ labelled by the displacement $\br\in\BR^d$
and the moment averages $\mu_\br(t)$ are the spatial correlation 
function $C(\br,t)$. The parametric model $P[\bX;\boalpha,t]$ is the distribution 
obtained by the {\it ansatz} $\phi({\bf x},t)=f(u({\bf x},t))$ where $u({\bf x},t)$ 
is a homogeneous Gaussian random field with mean zero and covariance 
$G(\br,t)=\langle u(\br,t)u(\bzed,t)\rangle$ and $f(z)$ is the stationary planar interface
solution of the TDGL equation (\ref{eq:TDGL}). In this case, it is the auxiliary 
correlation function $G(\br,t)$ which plays the role of the ``parameter'' 
$\alpha_\br(t).$ It is shown in \cite{Oono} for various cases how this  function
may be uniquely related to the ``moment'' $\mu_\br(t)=C(\br,t).$}.
In fact, perhaps the most familiar
application of the parametric approach is the use of the
Rayleigh-Ritz method in quantum mechanical calculations.  This is the
essential approach of our paper. 

\section{Equation-Free Computation}

Although we now have obtained a closed moment equation (equation
~\ref{eq:Veq}), we still need to determine the dynamical vector field
${\bf V}$. As explained above, this step can be a serious obstacle to
the practical implementation of PDF-based moment-closure (PDFMC).  
A method to calculate ${\bf V}$ is desirable that (i) does not require 
a radical revision each time the underlying code or model changes, 
and (ii) is relatively insensitive to the complexity of the PDFMC.
The equation-free approach of Kevrekidis and collaborators~\cite{EQNFREE} 
meets those requirements.  It permits one to work with much more sophisticated, 
physically realistic closures.

Equation-free computation is motivated by the simple observation that
numerical computations involving the closure equations ultimately do
not require closed formulae for the closure equations.  Instead, one
must only be able to sample an ensemble of system states $\bX$
distributed according to the closure {\it ansatz} $P(\bX,t;\boalpha)$
and then evolve each of these via equation (\ref{eq:Dynamics}) {\em
for short intervals of time}. Such sampling and subsequent dynamical
evolution would be necessary to calculate the statistics of interest
even when not using a closure strategy.  It is sufficient to have a (possibly
black-box) subroutine available which, given a specific state variable
$\bX(t)$ as input, returns the value of the state $\bX(t+\delta t)$
after a short time $\delta t$.  The ensemble of systems, each of which
satisfies equation (\ref{eq:Dynamics}), is evolved over a time
interval $\delta t$.  The moments/parameters ${\bomu}$ or $\boalpha$
are determined at the beginning and end of this interval and the time
derivative ${\dot {\bomu}}$ is estimated from the results of these
short ensemble runs.  This ``coarse timestepper'' can be used to
estimate locally the right hand side of the closure evolution
equations, namely ${\bf V}({\bomu}, t)$.

Coarse projective forward Euler (arguably the simplest of
equation-free algorithms) which we will use below illustrates the
approach succinctly: Starting from a set of coarse-grained initial
conditions specified by moments $\bomu (t)$ we first (a) {\em lift} 
to a consistent fine scale description, that is, sample the PDF {\it ansatz}
$P(\bX,t;\boalpha(t))$ to generate ensembles of initial conditions
$\bX$  for equation (\ref{eq:Dynamics}) consistent with the set $\bomu(t)$; 
(b) starting with these consistent initial conditions we evolve the fine scale 
description for a (relatively short) time $\delta t$; we subsequently {\em restrict} 
back to coarse observables by evaluating the moments $\bomu(t+\delta t)$ 
as ensemble-averages and (d) use the results to estimate locally the time
derivative $d\bomu/dt$. This is precisely the right hand-side
of the explicitly unavailable closure, obtained not through a closed
form formula, but rather through short, judicious computational
experiments with the original fine scale dynamics/code. Given 
this {\em local} estimate of the coarse-grained observable time
derivatives, we can now exploit the smoothness of their evolution in
time (in the form of Taylor series) and take a single long {\em
projective} forward Euler step:
\begin{eqnarray}
\bomu(t+\Delta t) = \bomu(t) + \Delta t\left[
\frac{\bomu(t+\delta t) - \bomu(t)}{\delta t}\right].
\end{eqnarray}
The procedure then repeats itself: lifting, fine scale evolution,
restriction, estimation, and then (connecting with continuum
traditional numerical analysis) a new forward Euler step. Beyond
coarse projective forward Euler, many other coarse initial-value
solvers (e.g. coarse projective Adams-Bashforth, and even implicit coarse
solvers) have been implemented; the stability and accuracy study of
such algorithms is progressing \cite{EQNFREE}.  These developments allow 
us to construct a {\em nonintrusive} implementation of PDF moment
closures, nonintrusive in the sense that we compute with the closures
without explicitly obtaining them, but rather by intelligently chosen
computational experiments with the original, fine-scale problem.

There is, however, an obvious objection to the equation-free 
implementation of moment-closures. Using the same ingredients,
one can clearly obtain an estimate of any statistics of interest
(for example, the moment-averages $\bomu(t)$) {\it without the need
of making any closure assumptions whatsoever.} This can be done
by the much simpler method of direct ensemble averaging. That is,
one can sample an ensemble of initial conditions $\bX$ from any 
chosen distribution $P_0(\bX)$, evolve each of these realizations 
according to the fine-scale dynamics of equation (\ref{eq:Dynamics}),
and then evaluate any statistics of interest at time $t$ by averaging 
over the ensemble of solutions $\bX(t).$ It would seem that this 
direct ensemble approach is much more straightforward and accurate
than the equation-free implementation of a moment-closure, which 
introduces additional statistical hypotheses. 

The response to this 
important objection is that the fine-scale dynamics (\ref{eq:Dynamics})
is often very stiff for the applications considered, in which the system contains
many-degrees-of-freedom interacting on a huge range of length- and 
time-scales. In contrast, the closure equation  (\ref{eq:Veq}) is much 
less stiff, because of statistical-averaging, and its solutions $\bomu(t)$ 
are much smoother in time (and space). Thus, to evolve an ensemble 
of solutions of the fine-scale dynamics  (\ref{eq:Dynamics}) from an 
initial time $t_0$ to a final time $t_0+T$ would require $O(T/\delta t)$ 
integration steps, where the time-step $\delta t$ is required to be very 
small by the  intrinsic stiffness of the micro-dynamics. In the closure
approach, the evolution of the moment equations (\ref{eq:Veq})
from time $t_0$ to time $t_0+T$ requires only $O(T/\Delta t)$ integration
steps, with (hopefully) $\Delta t\gg \delta t.$ Each of these closure integration
steps by an increment $\Delta t$ requires in the equation-free approach 
just one (or just a few) fine-scale integration step by an increment $\delta t.$ Thus, there
is an over-all savings by a (hopefully) large factor $O(\Delta t/\delta t).$ 
This crude estimate is based on a single step coarse projective
forward Euler algorithm; clearly, more sophisticated projective
integration algorithms can be used.

In all of them, however, the
computational savings are predicated on the smoothness of the closure
equations, and are governed by the ratio of the time that it takes to
obtain a good local estimate of $d\bomu/dt$ from full direct simulation
to the time that we can (linearly or even polynomially)
extrapolate $\bomu(t)$ in time.  It is also worth noting that a
variety of additional computational tasks, beyond projective
integration (e.g. accelerated fixed point computation) can be
performed within the equation-free framework

In the next section we show by a concrete example how significant
computational economy can be achieved with statistical moment 
closures implemented in the equation-free framework.

\section{A Numerical Example}

We illustrate here the equation-free implementation of moment-closures
for a canonical equation of phase-ordering kinetics~\cite{Bray}, the
stochastic time-dependent Ginzburg-Landau (TDGL) equation in one
spatial dimension.  This is written as
\begin{eqnarray} 
\frac{\partial \phi(x,t)}{\partial t} = D \Delta \phi(x,t) - V'(\phi(x,t))+ \eta(x,t)
\label{eq:TDGL}
\end{eqnarray}
where $\phi(x,t)$ represents a local order parameter, e.g. a magnetization.  
The noise has mean zero and covariance $\langle \eta(x,t) \eta(x',t')\rangle
=2kT\delta(x-x') \delta(t-t')$. The potential $V$ shall be chosen as 
$$ V(\phi) = \frac{1}{2}\phi^2 + \frac{1}{4}\phi^4 $$
to represent a single quartic/quadratic well. This stochastic dynamics has 
an invariant measure which is formally of Hamiltonian form $P_*[\phi]
\propto \exp(-H[\phi]/kT)$ where 
\be H[\phi] = \int \, [\frac{1}{2}D|\nabla\phi(x)|^2 + V(\phi(x))] \, dx. 
\lb{Ham} \ee
The Gibbsian measure $P_*[\phi]$ is approached at long times for any 
random distribution $P_0[\phi]$ of initial states. 

One of the simplest dynamical quantities of interest is the bulk 
magnetization $\overline{\phi}(t)=(1/V)\int \phi(x,t) dx,$ where $V$ 
is the total volume.  If the initial statistics are space-homogeneous,
then the ensemble average $\mu(t)=\langle\overline{\phi}(t)\rangle$
is also given by  $\mu(t)=\langle\phi(x,t)\rangle$ for any space point $x.$
Equation (\ref{eq:TDGL}) leads to a hierarchy of equations for 
statistical moments of $\phi(x,t)$. For example, the first moment 
satisfies the equation 
\begin{eqnarray} 
\frac{\partial \langle \phi(x,t) \rangle }{\partial t} = \Delta \langle
 \phi(x,t) \rangle - \langle \phi(x,t) \rangle - \langle \phi^3(x,t) \rangle.
\label{eq:FMOM}
\end{eqnarray}
The evolution of the mean total magnetization is thus a function of the
mean cubic total magnetization.  One could write a time evolution
equation for $\langle \phi^3 \rangle$, but it would involve a higher
order term $\langle\phi^5\rangle$, and so on. Each equation contains 
higher moments and therefore the hierarchy does not  close.

To close the equation for $\mu(t)$ we assume a parametric PDF  
of the form $P[\phi; \alpha]\propto \exp(-H[\phi;\alpha]/kT)$ where
$$ H[\phi;\alpha] = H[\phi] + \alpha \int \phi(x) \,dx
$$ is a perturbation of the Hamiltonian (\ref{Ham}) by a term
proportional to the moment variable $\xi[\phi]=(1/V)\int \phi(x)
\,dx.$ This is a special case of a general ``entropy-based'' closure
prescription proposed by Eyink and Levermore~\cite{EL}. This closure
scheme guarantees that $\alpha(t) \to 0$ at long times and therefore
the PDF {\it ansatz} $P[\phi;\alpha(t)]$ relaxes to the correct
stationary distribution $P_*[\phi]$ of the stochastic process.  The
determination of the parameter $\alpha$ given the moment $\mu$ is here
accomplished by Legendre transform
\begin{eqnarray}
 \alpha = {\rm argmax}_\alpha [\alpha \mu - F(\alpha) ],
\end{eqnarray}
where the ``moment-generating function'' $F(\alpha)=
\log \langle \exp[\alpha\int \phi(x)\,dx]\rangle_*$ and 
$\langle\cdot\rangle_*$ denotes average with respect to the invariant 
measure $P_*[\phi]$. 
The numerical optimization required for the Legendre transform is
well-suited to gradient descent algorithms such as the conjugate
gradient method, since
$$ (\partial/\partial \alpha)[\alpha \mu - F(\alpha) ]=\mu-\mu(\alpha), $$
where $\mu(\alpha)=\langle\xi\rangle_\alpha$ is the average of the 
moment-function in the PDF {\it ansatz} $P[\phi;\alpha].$ In simple 
cases, $F(\alpha)$ and $\mu(\alpha)=F'(\alpha)$ may be given
by closed analytical expressions. If not, then both of these averages 
may be determined together by Monte Carlo sampling techniques. 

In the numerical calculations below, we discretize equation (\ref{eq:TDGL})
using a forward Euler-Maruyama stochastic integrator and 3-point stencil 
for the Laplacian (other discretizations are possible).
\begin{eqnarray}
\phi(x, t+\delta t)= \phi(x, t) -\delta t [\phi(x,t) +
\phi^3(x,t)]+ \\ \nonumber \frac{D \delta t}{(\delta x)^2}
[\phi(x+\delta x, t) - 2\phi(x, t) + \phi(x-\delta x, t) ] + \\
\lb{TDGLDIS}
\nonumber \sqrt{2 kT(\delta t/\delta x) } N(x,t)
\end{eqnarray}
where $N(x,t)$ are independent, identically distributed standard
normal random variables for each space-time point $(x,t).$ The
invariant distribution of the stochastic dynamics space-discretized
in this manner has a Gibbsian form $\propto \exp(-H_\delta/ kT)$ with
discrete Hamiltonian
\begin{eqnarray}
H_\delta = \frac{D}{2\delta x}\sum_{\langle x,x' \rangle} (\phi(x)-\phi(x'))^2 \\ 
\lb{Ham-del} \nonumber + 
\sum_x \delta x [\frac{1}{2}\phi^2(x)+\frac{1}{4}\phi^4(x) ]
\end{eqnarray}
where $\langle x,x'\rangle$ are nearest-neighbor pairs. The closure {\it ansatz} 
can be adopted in the consistently discretized form $P_\delta[\phi; \alpha]\propto 
\exp(-H_\delta[\phi;\alpha]/kT)$ where
$$ H_\delta[\phi;\alpha] = H_\delta [\phi] + \alpha \sum_x \delta x\, \phi(x).  $$

In this numerical experiment, we integrate an $N=1000$ member ensemble of 
solutions of equation (\ref{TDGLDIS}), and measure the ensemble-averaged, 
global magnetization  $ \mu(t)=\langle \overline{\phi}(t) \rangle = 
(1/V)\sum_{x} \langle\phi(x,t) \rangle$ at each time-step. With this we compare 
the results of the entropy-based closure simulation 
implemented by the equation-free 
framework using also an ensemble with $N=1000$ samples. In this concrete
example, the projective integration scheme works as follows: Suppose we are
given the parameter $\alpha(t)$ at time $t$.  The mean $\mu(t)$ is first
calculated from the parametric ensemble at time $t$ by Monte Carlo sampling.
Next all $N$ samples are integrated over a short time-step  $\delta t$ to create a 
time-advanced ensemble. From this ensemble $\mu(t+\delta t)$ is calculated, 
which yields an estimate of the local time derivative.
$$  \dot{\mu}_{app}(t) = [\mu(t+\delta t)-\mu(t)]/\delta t.$$  
A large, projective Euler time-step of the moment average is then taken via
$$ \mu(t+\Delta t)=\mu(t) + \Delta t\,\dot{\mu}_{app}(t). $$ The parameter 
is finally updated by using the Legendre transform inversion to obtain 
$\alpha(t+\Delta t)$ from the known value $\mu(t+\Delta t).$ The cycle 
may now be repeated to integrate the closure equations by successive 
time-steps of length $\Delta t.$

A critical issue in general application of projective integration 
is the criterion to determine the projective time-step $\Delta t.$
For stiff problems with time-scale separation, the projective time step 
for stability purposes is of the order of (1/fastest ``slow group'' eigenvalues), 
while the ``preparatory'' simulation time is of the order of (1/slowest ``fast
group'' eigenvalue). Variants of the approach have been developed for
problems with several gaps in their spectrum~\cite{GK1}. Accuracy considerations
in real-time projective step selection can, in principle, be dealt
with in the traditional way for integrators with adaptive step-size
selection and error control: through on-line {\em a posteriori} error
estimates. An additional ``twist'' arises from the error inherent in the
estimation of the (unavailable) reduced time derivatives from the
ensemble simulations; issues of variance reduction and even on-line
hypothesis testing (are the data consistent with a local linear model?) 
must be considered. These are important research issues that are
currently explored by several research groups. Nevertheless, the main
factor in computational savings comes from the effective smoothness of
the unavailable closed equation: the separation of time scales between
the low-order statistics we follow and the higher order statistics
whose effect we model (and, eventually, the time scales of the
direct simulation of the original model).

Figure 1 is a plot comparing Projective Integration with Entropy
Closure and direct Ensemble Integration with equation ({\ref{eq:TDGL})
for diffusion constant $D=1000.0$ We have selected both the
``fine-scale'' integration step $\delta t$ and the ``coarse-scale''
projective integration step $\Delta t$ to be as large as possible,
consistent with stability and accuracy. Thus, only steps small enough
to avoid numerical blow-ups were considered. Then, values were
selected both for $\delta t$ and for $\Delta t$ so that the numerical
integrations with those time-steps differed by at most a few percent
from fully converged integrations with very small steps.  In this
manner, the time step required for the Euler-Maruyama integration of
({\ref{eq:TDGL}) was determined to be $ \delta t = 0.0004$. On the
other hand, for projective integration of the closure equation a time
step $\Delta t = 0.01$ could be taken. This indicates a gain in time
step by a factor of 25, which is also roughly the speed-up in the
algorithm or savings in CPU time. The present example is not as stiff
as equations that appear in more realistic applications, with a very
broad range of length- and time-scales, where even greater
computational economies might be expected.

\begin{figure}
\begin{center}
\includegraphics[width=3.0in]{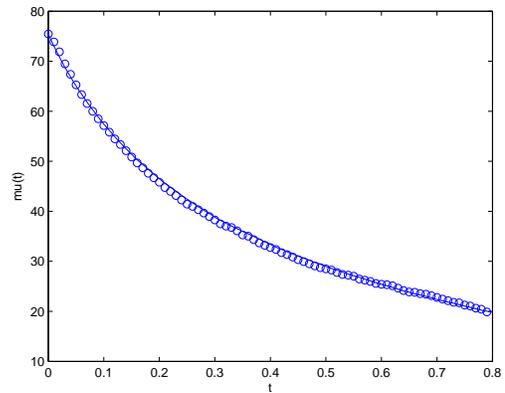}
\caption{ Mean total field as a function of time.  Line (symbols):
traditional (coarse projective) integration, respectively.  See the
text for a description of the stepsize selection.}
\end{center}
\end{figure}

\begin{figure}
\begin{center}
\includegraphics[width=3.0in]{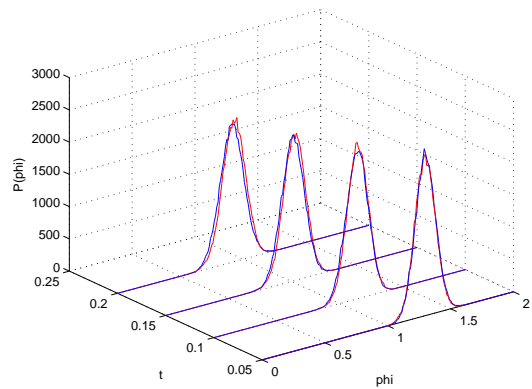}
\caption{ Comparison of the time dependent PDF's of the local field
$\phi(x,t)$ for the exact solution (blue) and for the projective integration
/ closure solution (red).}
\end{center}
\end{figure}

In general, the moment-closure results need not agree so well with
those of the direct ensemble approach, even when both are
converged. In the example presented here, there is good  agreement 
because the closure effectively captures the one-point PDF (see
Fig.2). This one-point PDF is the {\em only}} statistical quantity
that enters into Equation (\ref{eq:FMOM}) as long as the statistics are homogeneous and
the Laplacian term vanishes.

\section{Conclusions}

In this paper, we have described how one can combine recently developed
equation-free methods with statistical moment closures to model nonlinear 
problems.  With this method we can numerically integrate complex nonlinear
systems, for which closure equations may not be available in closed form.  
In the example presented here the specific entropy-based closure we 
selected has an H-theorem which guarantees relaxation to the equilibrium 
state of the original dissipative dynamics.  However, we stress that the 
general approach outlined above can be used with a variety of closure methods.

The equation-free method has the potential to
enhance the flexibility, power, and applications set of the statistical
moment closure approach.  Since little or no analytic work is
required, the sophistication of statistical moment closures can 
greatly enhanced beyond Gaussian PDF {\it ans\"atze}. The ``practical usefulness''
criterion 
for parametric PDF models that they permit analytical calculations is 
replaced by the criterion that they can be efficiently sampled.
We believe that this approach can significantly increase  
the usefulness of closure methods.

In order to model systems like global climate, oceans, and reaction
diffusion processes in systems biology, one will have to construct
more complex closures.  These will likely include higher order
moments, correlation functions of the relevant variables, highly
non-Gaussian statistics, etc.  As the closures become more complex,
the lifting step will require more efficient sampling approaches.  One
will likely have to use nonlocal, accelerated sampling methods.  One
will also likely employ the latest in adaptive time and adaptive mesh
methods to optimize performance for large-scale problems.

\section{Acknowledgements}
This work, LA-UR-07-2218, was carried out in part at Los Alamos
National Laboratory under the auspices of the US National Nuclear
Security Administration of the US Department of Energy. It was
supported under contract number DE-AC52-06NA25396. The work of IGK was
partially supported by DARPA and by and US DOE(CMPD). G. Eyink was
supported by NSF-ITR grant, DMS-0113649.

\end{document}